\documentclass[a4paper,11pt]{article}
\usepackage{pos}

\title{Wind parameters of the new LBV in NGC1156}
\author*[a]{Y. Solovyeva}
\author[a]{A. Kostenkov}
\author[a]{E. Dedov}
\author[a]{A. Vinokurov}

\affiliation[a]{Special Astrophysical Observatory,\\
Nizhnij Arkhyz 369167, Russia}

\emailAdd{solovyeva@sao.ru}
\emailAdd{kostenkov@sao.ru}

\abstract{%We model the spectrum of a new luminous blue variable (LBV) in the NGC1156 using the non-LTE code CMFGEN intended for calculations of the extended atmospheres of massive stars with mass outflow. 
In this work we present the spectrum modeling results for the newly discovered luminous blue variable (LBV) in the NGC\,1156 galaxy. Extended atmosphere models were calculated using the non-LTE code CMFGEN. 
We have obtained the luminosity of the discovered LBV $L\simeq (1.6\pm0.2) \times 10^{6}\, L_{\odot}$, effective temperature $T_{\text{eff}}=7.9\pm0.4$\,kK and mass-loss rate $\dot{M}f^{-0.5}= (8.2\pm1.0) \times 10^{-4}\,M_{\odot}\,\text{yr}^{-1}$. The hydrogen abundance in the wind is $\approx20\,$\% for the metallicity $Z=0.5\,Z_{\odot}$ of the host galaxy.}%with overabundance of nitrogen ($X_\text{N}/ X_{\odot} = 5.41$) and underabundance of carbon ($X_\text{C}/ X_{\odot} = 0.20$).}
\FullConference{%
  The Multifaceted Universe: Theory and Observations -  2022 (MUTO2022)\\
  23-27 May 2022\\
  SAO RAS, Nizhny Arkhyz, Russia\\}

\begin{document}
\maketitle

\section{Introduction}
Luminous blue variables are a rare type of luminous massive stars (M>25 M$\odot$, L$_{Bol}\sim 10^6$\,L$\odot$, \cite{Humphreys2016}) at an advanced stage of evolution. They are characterized by noticeable photometric and spectral variability, and their important observational feature is the S\,Dor type variability, which consists of stellar brightness variations up to $\approx 2 - 2.5^m$ simultaneously with spectral changes at an approximately constant (within 0.2 dex) bolometric luminosity \cite{Groh2009}.

We discovered two new LBVs in the galaxy NGC\,1156 (D=7.0 Mpc, \cite{Solovyeva2022}). The more luminous of the two, J025941.21+251412.2, showed brightness variations with an amplitude of $\Delta \text{R}_c = 0.84 \pm 0.23^m$ \cite{Solovyeva2022}. Here we present the results of modeling its spectrum at maximum visual brightness using the CMFGEN code \cite{Hillier1998}.

\section{Methods}
For modeling, we chose a spectrum with the best resolution and the highest signal-to-noise ratio, which was obtained using BTA/SCORPIO-2 in the long slit mode (see details in \cite{Solovyeva2022}). The observed spectrum has a spectral resolution of $\sim 4.5$\AA{} in the 3600-7300\,\AA{} spectral range. 

The spectrum of J025941.21+251412.2 contains hydrogen lines with broad wings and numerous Fe\,II and [Fe\,II] lines, which are characteristic of many types of massive stars with a strong gas outflow. In addition to the lines of the object itself, the spectrum exhibits narrow components of hydrogen lines, [O\,III]\,$\lambda\lambda4959,5007$, [N\,II]\,$\lambda\lambda6548,6583$, [S\,II]\,$\lambda\lambda6716,6731$ and some other emissions of the nebula surrounding the object together with the neighboring groups of young stars. To remove the nebula contribution, we extracted its pure spectrum in a star-free region located in the slit near the object, then multiplied this spectrum by a scaling factor and subtracted it from the observed spectrum of the object. The scaling factor was calculated in such a way as to achieve the most complete subtraction of all nebula lines from the object spectrum.

Numerous Fe\,II lines in the spectrum probably indicates the low ionization state of the wind. In addition, the absence of helium lines in the spectrum may also indicate a relatively low photospheric temperature ($\lesssim$11\,kK) during the corresponding observation period. 

Initially, we selected several "cold" models ($T_{\text{eff}}\approx$11-12\,kK) from a pre-calculated grid of extended model atmospheres. After that, we scaled the mass-loss rate, radius and luminosity of the most suitable model according to the photometric data, V=$19.41$\,mag. The best-fit model was convolved with the absorption curve \cite{Fitzpatrick1999} using interstellar reddening $A_{\text{v}}=0.9$ obtained from the Balmer decrement of the star's surrounding nebula.

\begin{figure*}[h!]
\centering
    \includegraphics[width=0.98\linewidth]{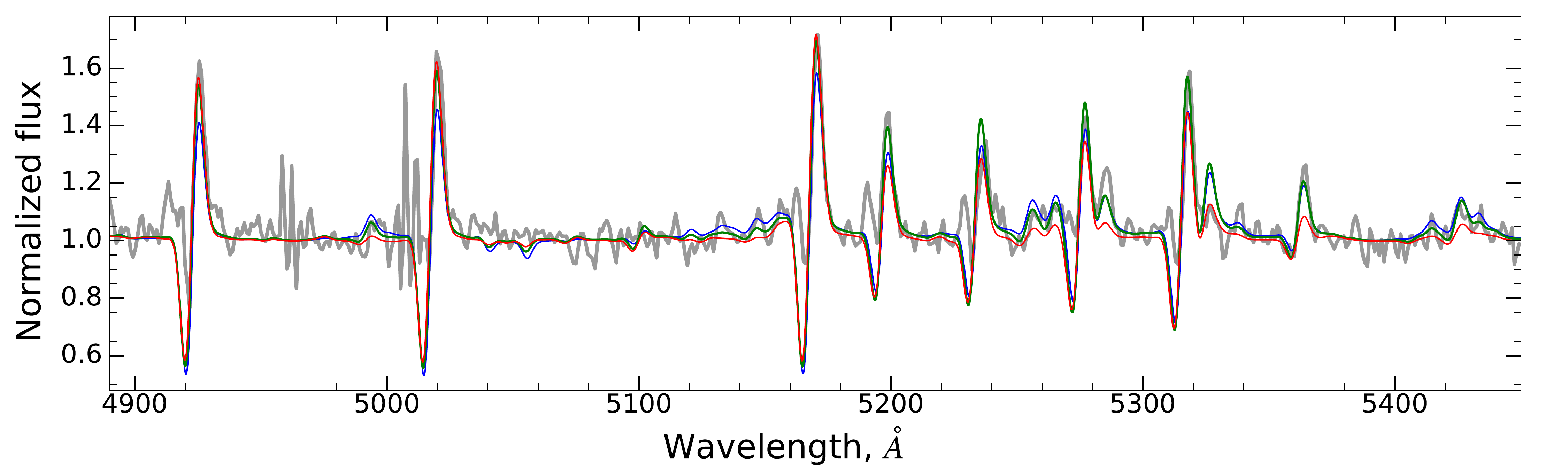}
    \caption{Comparison of the Fe\,II $\lambda$4924, $\lambda$5018, $\lambda$5169, $\lambda$5197-5427 lines in best-fit models with different effective temperatures: 7.3\,kK (blue solid line), 7.9\,kK (green solid line) and 8.4\,kK (red solid line). The observed spectrum of LBV J025941.21+251412.2 is marked by a grey solid line.}

    \label{ngc1156_diff_temps}
\end{figure*}

There are no direct temperature indicators (e.g. He\,I\,/\,He\,II lines ratio) in the optical spectrum, and therefore, we used different Fe\,II lines for effective temperature estimations. We built a set of models with different temperatures in the range of $\sim$\,7-11\,kK. Fig.\,\ref{ngc1156_diff_temps} shows several model spectra with different temperatures and a fixed photospheric radius $R_{2/3}\approx680\,R_{\odot}$ in a spectral range covering several Fe\,II lines. 
As seen in Fig\,\ref{ngc1156_diff_temps}, the ratios of the Fe\,II $\lambda$4924, $\lambda$5018, $\lambda$5169 line equivalent widths to those of the Fe\,II $\lambda$5197-5427 line series decrease with lower temperatures. The chosen line ratios are highly sensitive to temperature changes. Hence, we obtained relatively accurate photospheric temperature estimates. The effective temperature ranges from 7.5 to 8.3\,kK in models that showed the best agreement with the observed spectrum in the Fe\,II lines. The model absorption component of several Fe\,II lines is deeper than in the observed spectrum, probably due to the presence of forbidden [Fe\,II] lines in the blue wings of the Fe\,II lines formed in the previously ejected low density gas. 

We obtained the mass-loss rate by fitting strong Balmer series lines. The hydrogen abundance estimate 20\,\% by mass fraction was based on the equivalent width ratio of the $H_{\beta}$, $H_{\gamma}$, $H_{\delta}$ lines to Fe\,II $\lambda$5169. Hydrogen abundance values range from 15 to 20\,\% in the most suitable models.

The metallicity $Z=0.5\,Z_{\odot}$ of the host galaxy NGC\,1156 \cite{Kim2012} was used in all models. The nitrogen and carbon abundances were assumed to be $X_\text{N}/ X_{\odot} = 5.41$ and $X_\text{C}/ X_{\odot} = 0.20$, respectively.

There are no forbidden wind lines with clearly identified profiles (e.g. [N\,II] $\lambda$5755) in the optical spectrum. Thus, the terminal wind velocity cannot be measured directly from the spectrum, and therefore, we used only the widths of strong Balmer series emission lines for crude wind velocity estimates. We assumed terminal velocity $V_{\infty}=300\,\text{km\;s}^{-1}$ and a simple velocity law $\beta=4.0$ \cite{Hillier1989} in the best-fit model. The microclumping in the wind cannot be estimated properly due to the strong wide wings in the profiles of the $H_{\alpha}$, $H_{\beta}$ lines. Thus, a typical LBV volume filling-factor $f=0.3$ was chosen for calculations (e.g. \cite{Groh2009}).

\section{Results}

\begin{figure*}[h!]
\centering
    \includegraphics[width=0.95\linewidth]{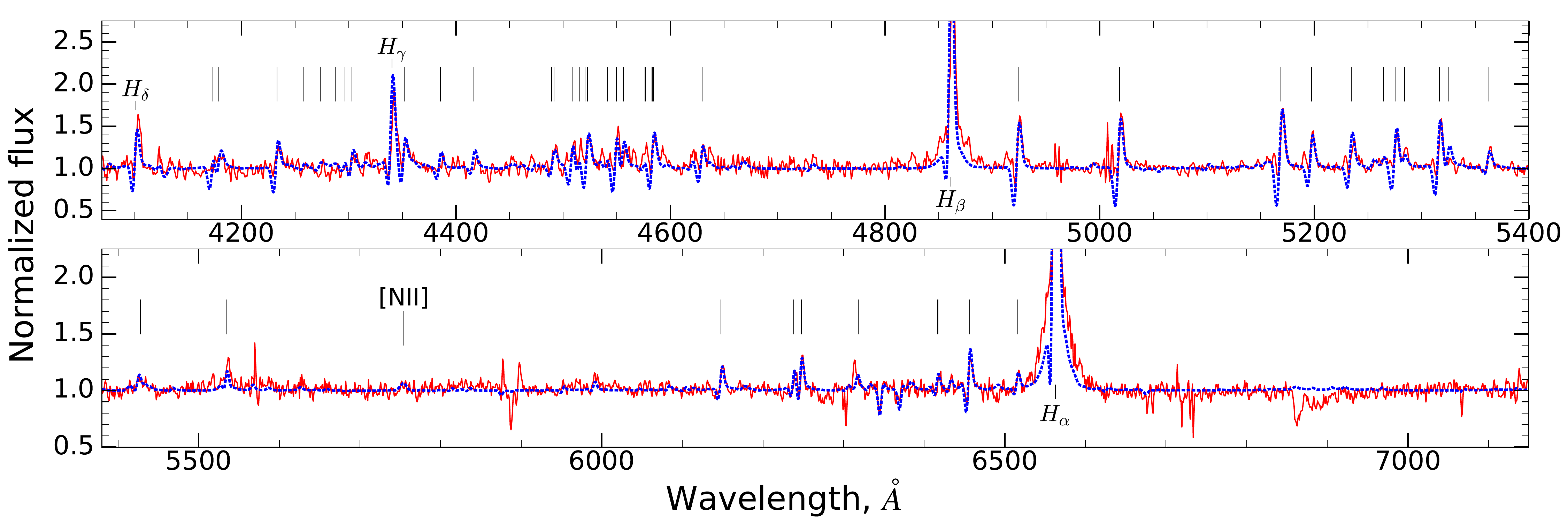}
    \caption{The best-fit model spectrum (blue dashed line) compared with the observed spectrum of LBV J025941.21+251412.2 (red solid line). Fe\,II lines are marked with black vertical lines above the spectrum.}

    \label{ngc1156_lbv_model_spec}
\end{figure*}

The normalized optical spectra and best-fit model of LBV J025941.21+251412.2 with the most important lines are shown in Fig.\,\ref{ngc1156_lbv_model_spec}.

The luminosity of LBV J025941.21+251412.2 is $L\simeq (1.6\pm0.2) \times 10^{6}\, L_{\odot}$, the effective temperature is $T_{\text{eff}}=7.9\pm0.4$\,kK and the mass-loss rate is $\dot{M}f^{-0.5}= (8.2\pm1.0) \times 10^{-4}\,M_{\odot}\,\text{yr}^{-1}$. The photospheric radius is $R_{2/3}\simeq680\,R_{\odot}$ for the photometric data, distance and interstellar reddening presented above.

Hydrogen abundance in the wind of J025941.21+251412.2 is significantly lower ($\sim20\,\%$) than the one for studied LBVs ($\sim30$-40\,\%, \cite{Groh2009}). 
The chemical composition of J025941.21+251412.2 is similar to  that of a late-WN star, which probably indicates the advanced evolutionary stage of the discovered star.

\section*{Acknowledgements}

Modeling was performed as part of the government contract of SAO RAS approved by the Ministry of Science and Higher Education of the Russian Federation. Observations with the SAO RAS telescopes are supported by the Ministry of Science and Higher Education of the Russian Federation. The renovation of telescope equipment is currently provided within the national project ''Science and universities''. 

\bibliographystyle{JHEP}
\bibliography{biblio.bib}

%\begin{thebibliography}{99}

%\bibitem{Humphreys2016} Humphreys R.~M., Weis K., Davidson K., Gordon M.~S., Astrophys. J. 825 (2016) 64. %doi:10.3847/0004-637X/825/1/64;

%\bibitem{Groh2009} Groh J.~H., Hillier D.~J., Damineli A., Whitelock P.~A., Marang F., Rossi C., 2009, ApJ, 698, 1698. %doi:10.1088/0004-637X/698/2/1698

%\bibitem{Solovyeva2022} Solovyeva Y., Vinokurov A., Tikhonov N., Kostenkov A., Atapin K., Sarkisyan A., Moiseev A., et al., 2022, paper submitted for publication in MNRAS, arXiv:2208.05858 %in preparation 

%\bibitem{Hillier1998} Hillier D.~J., Miller D.~L., 1998, ApJ, 496, 407. %doi:10.1086/305350

%\bibitem{Afanasiev2011} Afanasiev V.~L., Moiseev A.~V., 2011, BaltA, 20, 363. %doi:10.1515/astro-2017-0305

%\bibitem[\protect\citeauthoryear{Fitzpatrick}{1999}]{Fitzpatrick1999} Fitzpatrick E.~L., 1999, PASP, 111, 63. 
%doi:10.1086/316293

%\bibitem{Kim2012} Kim S.~C., Park H.~S., Kyeong J., Lee J.~H., Ree C.~H., Kim M., 2012, PASJ, 64, 23. 
%doi:10.1093/pasj/64.2.23

%\bibitem[\protect\citeauthoryear{Hillier}{1989}]{Hillier1989} Hillier D.~J., 1989, ApJ, 347, 392. 
%doi:10.1086/168127

%\bibitem[\protect\citeauthoryear{Najarro}{2001}]{Najarro2001} Najarro F., 2001, ASPC, 233, 133

%\end{thebibliography}

\end{document}